\documentclass{article}

\usepackage{microtype}
\usepackage{graphicx}
\usepackage{subcaption}
\usepackage{booktabs}
\usepackage{multirow}
\usepackage[table]{xcolor}
\usepackage{placeins}
\definecolor{firstBest}{rgb}{0.86, 1, 0.86}
\definecolor{COLOR_MEAN}{HTML}{f0f0f0}

\usepackage{hyperref}

\usepackage[preprint]{paperstyle}

\usepackage{amsmath}
\usepackage{amssymb}
\usepackage{mathtools}
\usepackage{amsthm}
\usepackage{natbib}

\usepackage[capitalize,noabbrev]{cleveref}

\theoremstyle{plain}

\theoremstyle{definition}

\theoremstyle{remark}

\papertitlerunning{VLD-RAG: Agentic Vision--Language RAG for Long Visually-Rich Documents}

\begin{document}

\twocolumn[
  \papertitle{VLD-RAG: Agentic Vision--Language Retrieval-Augmented Generation for Long, Visually-Rich Multi-Page Documents}

  \begin{paperauthorlist}
    \paperauthor{Seonok Kim}{mazelone}
  \end{paperauthorlist}

  \paperaffiliation{mazelone}{Mazelone}

  \papercorrespondingauthor{Seonok Kim}{seonokrkim@gmail.com}

  \paperkeywords{Multimodal RAG, Visual Document Retrieval, Visually-Rich Documents, Long-Context Documents, Multi-Page Document Understanding, Hybrid Retrieval, Agentic RAG, Evidence Verification, LongDocURL, MMLongBench-Doc}

  \vskip 0.3in
]

\printAffiliationsAndNotice{}

\begin{abstract}
  Visually-rich documents such as reports, slides, and manuals often distribute the evidence needed to answer a question across multiple pages, mixing text with layout cues, tables, charts, and figures. This work studies multimodal retrieval-augmented generation for question answering over such visually-rich long documents, where retrieval must select evidence pages that include both textual and visual signals. We present VLD-RAG, an agentic multimodal RAG framework for multi-page evidence retrieval and cross-page reasoning over long documents. VLD-RAG builds a page-preserving multimodal index that stores parsed text, page-level metadata, and dense visual representations, and uses a hybrid retrieval strategy that combines keyword-based sparse search with dense semantic queries to identify candidate sources and evidence pages. A verifier-guided agent workflow coordinates a Retrieval Agent, Answer Agent, and Validation Agent to broaden evidence coverage, detect missing citations, and refine retrieval requests when needed. We evaluate retrieval with Top-1 and Top-5 evidence-page accuracy and generation with generalized accuracy, and show that VLD-RAG improves both evidence-page retrieval and end-task question answering on visually-rich long-document benchmarks, including LongDocURL and MMLongBench-Doc, outperforming previous vision-based retrieval baselines. These findings highlight that coordinated agent verification and multimodal hybrid retrieval are crucial for reliable grounding when correct answers depend on evidence scattered across pages.

\end{abstract}

\section{Introduction}
\label{sec:introduction}

Multimodal document understanding has become increasingly important as real-world documents frequently combine natural language with rich visual structure, including tables, charts, diagrams, and complex layouts. While retrieval-augmented generation (RAG) has improved knowledge-grounded question answering in text-centric settings, standard RAG pipelines often degrade on visually rich documents: converting pages to plain text can be lossy, obscuring layout semantics, spatial relationships, and non-textual cues that are essential for correct interpretation. As a result, both retrieval and generation may fail to ground answers in the appropriate evidence, especially when the required information is distributed across multiple pages.

\begin{figure*}[t]
    \centering
    \includegraphics[width=\textwidth]{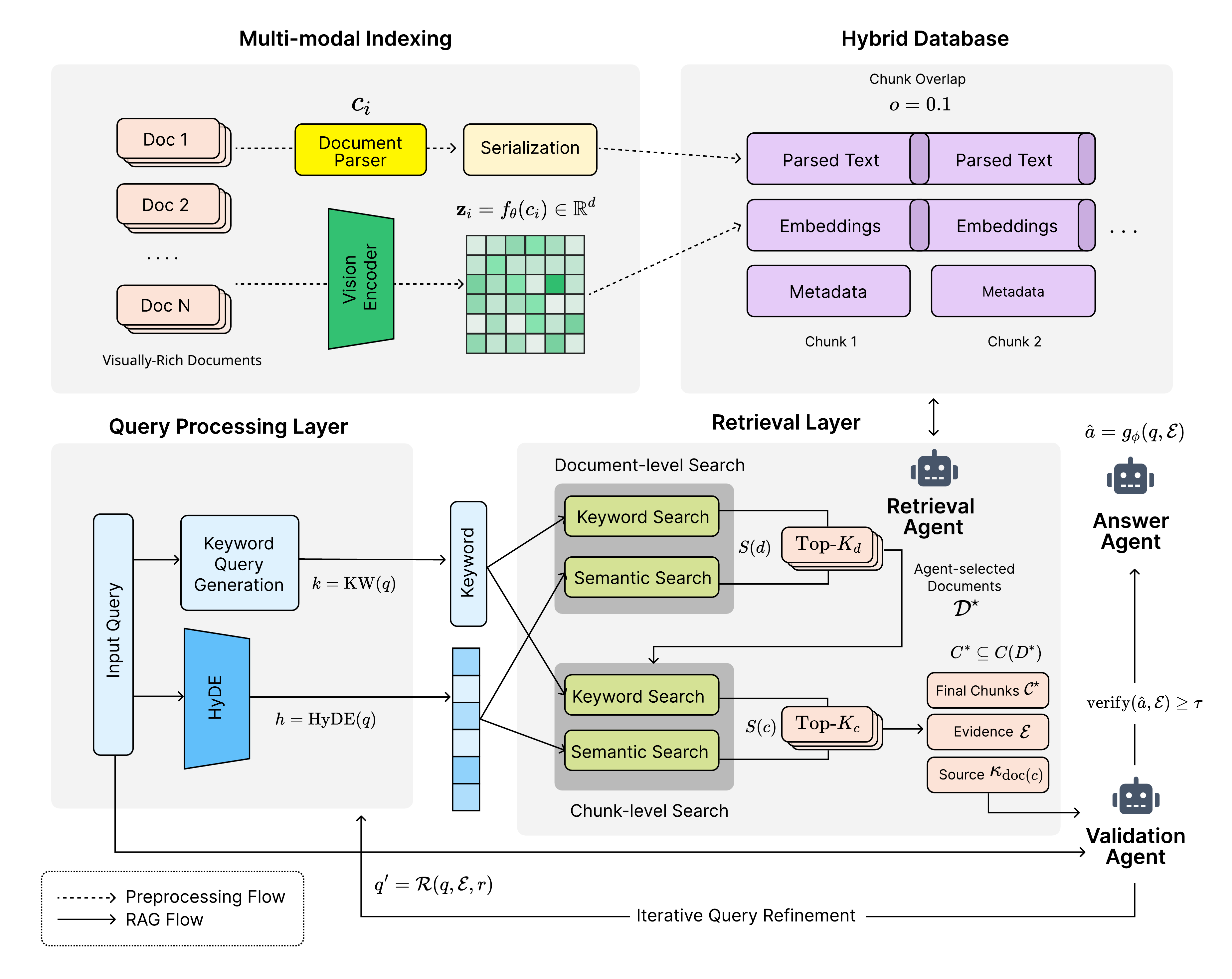}
    \caption{Overview of the VLD-RAG framework. The pipeline processes a user query through multimodal query decomposition to extract retrieval intents, then performs hybrid visual--textual retrieval with modality-consistent fusion and optional iterative refinement. Retrieved candidates are passed to the generator for evidence-packed answer production; validation can trigger query refinement when evidence is missing or confidence is low.}
    \label{fig:overview}
\end{figure*}

Recent advances in vision-language models have enabled new approaches to document understanding that operate directly over page images. Vision-first document RAG lines of work \cite{yu2025visrag} suggest that retrieving by page images can better preserve visual information than purely text-based parsing. Nevertheless, existing approaches still face key challenges. First, many systems depend heavily on extracted text or OCR artifacts, which can miss structural signals and introduce noise in visually dense regions. Second, effectively combining lexical signals (e.g., identifiers, rare entities, exact numbers) with multimodal semantic similarity remains non-trivial, and naive score fusion can be brittle when modalities disagree. Third, long documents often require multi-step evidence discovery, where the answer is supported by multiple scattered pieces of information across pages, making single-shot retrieval insufficient.

To address these limitations, we propose VLD-RAG: a Vision--Language Document Retrieval-Augmented Generation framework that performs iterative multimodal querying for reliable multi-page evidence retrieval and cross-page reasoning. VLD-RAG introduces two core design principles. (\textit{i}) We adopt a dual-index architecture that preserves complementary signals: a visual index of page images represented by dense vision-language embeddings, and a text index built from structured markdown serialization of the pages. The markdown representation preserves document structure (e.g., tables, figures, headings, and region boundaries) and supports faithful reconstruction of evidence during generation. (\textit{ii}) We introduce modality-consistent fusion coupled with an agentic retrieval loop. Rather than treating sparse and dense retrieval as interchangeable, VLD-RAG explicitly models when the two modalities should agree and uses verification-driven refinement to recover missing evidence, building on insights from iterative visual document RAG \cite{wang2025vidorag} and consistency-aware multimodal QA \cite{suri2024visdommultidocumentqavisually}.

Concretely, VLD-RAG operates in three stages. First, it performs multimodal query decomposition to produce structured retrieval intents, including salient entities, expected evidence types (e.g., table/chart/paragraph), and optional visual anchors. Second, it conducts hybrid visual--textual retrieval by combining sparse lexical retrieval with dense multimodal retrieval, followed by consistency-constrained fusion that down-weights candidates exhibiting cross-modal disagreement. When the fused evidence set is insufficient, a verifier provides refinement signals that trigger additional retrieval with updated intents. Third, VLD-RAG performs evidence-grounded generation, packing page-level evidence (and optional regions/snippets) into a structured context that maximizes grounding while controlling context length, and validating the produced answer against the retrieved evidence.

We evaluate VLD-RAG on visually rich long-document benchmarks, including LongDocURL and MMLongBench-Doc, and observe consistent gains in both evidence retrieval and end-task question answering relative to vision-based retrieval baselines and hybrid alternatives. In summary, our contributions are:
(1) a vision--language document RAG framework with a dual-index representation that preserves both visual layout cues and precise lexical signals via structured markdown serialization;
(2) a modality-consistent hybrid retrieval strategy that improves robustness under cross-modal mismatch; and
(3) a verifier-guided agentic retrieval loop that enables reliable multi-page evidence discovery for cross-page document question answering.

\section{Related Work}
\label{sec:related_work}

\subsection{Cross Modality Knowledge Base}

Graph-based approaches have shown effectiveness in knowledge retrieval and reasoning for multimodal systems. \citet{bu-etal-2025-query} proposed Query-Driven Multimodal GraphRAG for dynamic local knowledge graph construction tailored to query semantics. \citet{han2025retrievalaugmentedgenerationgraphsgraphrag} explored GraphRAG for retrieval-augmented generation with graphs, while \citet{wang2025taming} introduced a graph-based knowledge retrieval and reasoning approach for MLLMs to conquer unknown domains. \citet{lu2025karma} leveraged multi-agent LLMs for automated knowledge graph enrichment. \citet{hong2025knowledgebasedvisualquestionanswer} developed a knowledge-based visual question answering system with multimodal processing, retrieval and filtering. \citet{lin-etal-2025-makar} proposed MAKAR, a multi-agent framework based knowledge-augmented reasoning for grounded multimodal named entity recognition.

\subsection{Multi-modal Document Retrieval}

Cross-modal retrieval has been a key focus in multimodal document understanding. \citet{fang-etal-2025-cart} proposed CART, a generative cross-modal retrieval framework with coarse-to-fine semantic modeling that treats generating identifiers as retrieval targets. \citet{Zhang_2025_CVPR} introduced a method for improving universal multimodal retrieval by multimodal large language models, bridging modalities effectively. \citet{sun-etal-2025-unveil} presented Unveil, a unified visual-textual integration and distillation approach for multi-modal document retrieval. Recent document understanding models have shown significant improvements: \citet{hu-etal-2025-mplug} introduced mPLUG-DocOwl2 for high-resolution compressing for OCR-free multi-page document understanding, while \citet{ye2025mplugowl} presented mPLUG-Owl3 for long image-sequence understanding in multi-modal large language models. \citet{zhu2025inteexpladva} explored advanced training and test-time recipes for open-source multimodal models with InternVL3. \citet{zhang2024ocr} evaluated the cascading impact of OCR on retrieval-augmented generation, showing that OCR can hinder RAG performance in certain scenarios. \citet{deng2025longdocurl} introduced LongDocURL, a comprehensive multimodal long document benchmark integrating understanding, reasoning, and locating.

\subsection{Retrieval-Augmented Generation for Multimodal Documents}

Recent advances in retrieval-augmented generation (RAG) have shown strong promise for multimodal document understanding. \citet{tanaka2025vdocrag} introduced VDocRAG for visually-rich documents, and \citet{chen2025svrag} proposed SV-RAG, which applies LoRA-based contextual adaptation of MLLMs for long-document understanding. \citet{yu2025visrag} presented VisRAG, a vision-based RAG system for multi-modality documents, and \citet{sun2025visrag20evidenceguidedmultiimage} extended it with VisRAG 2.0 by incorporating evidence-guided multi-image reasoning.

Several works further study multimodal RAG for long or multi-document settings. \citet{shi2026urag} proposed URaG, a unified retrieval and generation approach for long document understanding in multimodal LLMs. \citet{suri2024visdommultidocumentqavisually} developed VisDoM for multi-document QA with visually rich elements, while \citet{Cho2024M3DocRAG} introduced M3DocRAG, highlighting the importance of multi-modal retrieval for multi-page, multi-document understanding. \citet{wu2025molorag} proposed MoLoRAG, which bootstraps document understanding via logic-aware retrieval, demonstrating the benefit of structured retrieval signals in multimodal documents.

In parallel, comprehensive benchmarks and analyses have emerged to evaluate multimodal RAG systems. \citet{li2024benchmarkingmultimodalretrievalaugmented} introduced a benchmarking framework based on dynamic VQA, and \citet{hu2024mragbench} presented MRAG-Bench, a vision-centric evaluation for retrieval-augmented multimodal models. \citet{chan2025mmuragent} developed MMU-RAGent, a large-scale user-centric benchmark. Domain-focused resources include FinRAGBench-V for financial RAG with visual citation \citep{zhao2025finragbench} and MMed-RAG for medical vision-language models \citep{xia2024mmedrag}. Beyond capability, \citet{hao2025rap} studied retrieval-augmented personalization (RAP), and \citet{zhang2025poisonedeye} investigated knowledge poisoning attacks, highlighting security risks in retrieval-augmented vision-language systems. However, existing systems often struggle with reliable multi-page evidence retrieval when information is distributed across pages, and few approaches explicitly model cross-modal consistency or provide verifier-guided refinement mechanisms for improving retrieval robustness.

\subsection{Vision Language RAG with Agents}

Agent-based approaches have enhanced vision-language RAG systems with dynamic reasoning capabilities. \citet{wang2025vidorag} introduced ViDoRAG, which uses dynamic iterative reasoning agents for visual document retrieval-augmented generation. \citet{liu2025cadencerag} proposed CadenceRAG for context-aware and dependency-enhanced retrieval augmented generation for holistic video understanding. \citet{biswas2025raven} introduced RAVEN, a query-guided representation alignment approach for question answering over audio, video, embedded sensors, and natural language. \citet{blau2024gramglobalreasoningmultipage} developed GRAM for global reasoning in multi-page VQA. \citet{han2025mdocagent} proposed MDocAgent, a multi-modal multi-agent framework for document understanding that leverages multiple agents for comprehensive document analysis. \citet{su2025skvqasynthet} introduced SK-VQA for synthetic knowledge generation at scale for training context-augmented multimodal LLMs. \citet{wangretrieval} proposed retrieval-augmented perception, combining high-resolution image perception with visual RAG.

\section{Method}
\label{sec:method}

\subsection{Task Definition}

We study \textbf{Vision--Language Document Retrieval-Augmented Generation (VLD-RAG)}: given a \textbf{visually rich document} $D$ (a set of page images $\{p_i\}_{i=1}^{N}$, optionally with parsed and serialized content) and a user query $q$, the model produces an answer $y$ while retrieving a small set of supporting visual/textual evidence $E \subset D$.

Unlike text-only RAG, VLD-RAG must preserve and exploit \textbf{layout, tables, charts, and non-textual visual cues}, where parsing to text can be lossy (as also highlighted in vision-first document RAG lines of work \cite{yu2025visrag}).

\subsection{Overview of the VLD-RAG Framework}

VLD-RAG is a \textbf{three-stage pipeline}: (1) \textbf{Multimodal Processing \& Querying} produces structured retrieval intents (entities, visual anchors, sub-questions); (2) \textbf{Hybrid Visual--Textual Retrieval with Consistent Fusion} retrieves candidate pages/regions using both visual and text signals, then fuses scores with modality-consistency constraints; (3) \textbf{Evidence-Packed Generation} produces an answer grounded in the retrieved page regions/snippets. This design is inspired by (i) \textbf{iterative reasoning / agentic retrieval} for visual documents \cite{wang2025vidorag}, and (ii) \textbf{consistency-constrained modality fusion} in multi-document visually rich QA \cite{suri2024visdommultidocumentqavisually}.

\subsection{Dual-Space Document Representation}

For each page image $p_i$, we compute a dense embedding (\textbf{visual index}, vision embeddings):
\begin{equation}
\mathbf{v}_i = f_{\theta}(p_i)
\end{equation}
where $f_{\theta}$ (equivalently denoted as $f_{\text{vis}}$) is a vision-language retriever encoder parameterized by $\theta$ that captures both visual layout and textual semantics in a unified representation space. This follows the ``retrieve by page image'' philosophy in vision-first document RAG \cite{yu2025visrag}, where embeddings $\mathbf{v}_i \in \mathbb{R}^d$ preserve multimodal information. We also parse pages into structured markdown format, preserving layout information including tables, figures, and text regions. These parsed markdown representations are serialized and indexed as chunk embeddings $\mathbf{t}_{i,j}$ (\textbf{text index}). The dual use of vision embeddings and text resources enables retrieval that leverages both visual semantics and precise text matching (IDs, numbers, rare entities) as well as structural information from markdown.

\subsection{Multimodal Query Formulation}

Given $(q, D)$, VLD-RAG first produces \textbf{structured retrieval intents}:

\begin{equation}
\begin{split}
\mathcal{I} = \{&\text{entities}, \text{visual anchors}, \\
&\text{sub-queries}, \text{expected evidence type}\}
\end{split}
\end{equation}

The model extracts key entities (e.g., component names, regulation IDs, table headers) and visual anchors (e.g., ``the chart titled \ldots'', ``the table in Section \ldots''), which improves downstream retrieval robustness---consistent with prior observations that query quality is a bottleneck in knowledge-grounded VQA/RAG \cite{hong2025knowledgebasedvisualquestionanswer}.

\subsection{Modality-Consistent Multimodal Retrieval}

Our hybrid retrieval framework integrates sparse lexical retrieval with dense vision-language semantic retrieval. This design addresses the observation that evidence pages in long documents often differ substantially from user queries in surface form, particularly when answers are embedded within complex layouts, tables, or figures.

We first extract keywords from the query using a keyword extraction function (\textbf{sparse lexical retrieval}):
\begin{equation}
k = \text{KW}(q)
\end{equation}
where $\text{KW}(\cdot)$ extracts key terms, entities, and phrases from the query. We then retrieve an initial set of candidate pages using sparse lexical retrieval with standard parameters (term frequency saturation $k_1=1.5$, length normalization $b=0.75$), capturing exact term and phrase matches:
\begin{equation}
\mathcal{P}_{\text{sparse}} = \text{SparseRetrieval}(k, D, K)
\end{equation}
In parallel, we generate a hypothetical document embedding for dense semantic retrieval:
\begin{equation}
h = \text{GenerateHypotheticalDoc}(q)
\end{equation}
where a language model (Gemma3n 4B \citep{gemma_3n_2025}) generates a hypothetical document $h$ conditioned on the input question. This synthetic document approximates the content of an ideal answer-bearing page and bridges the semantic gap between questions and document pages. The hypothetical document is then encoded into a dense vision-language representation:
\begin{equation}
\mathbf{e}_h = f_{\text{vis}}(h)
\end{equation}
where $f_{\text{vis}}$ is the ColPali vision encoder. Dense retrieval is performed by matching this representation against document page embeddings:

\begin{equation}
\mathcal{P}_{\text{dense}} = \text{ANN-Search}(\mathbf{e}_h, \{\mathbf{v}_i\}, K)
\end{equation}
We retrieve top-$K$ candidates from both branches (\textbf{candidate generation}): sparse retrieval yields $\mathcal{P}_{\text{sparse}}$ from keyword-based search, and dense retrieval yields $\mathcal{P}_{\text{dense}}$ from vision embedding similarity search. For higher precision, we optionally rerank candidates using an MLLM-based retriever/reranker design aligned with universal multimodal retrieval (UMR) insights (strong text$\leftrightarrow$image$\leftrightarrow$mixed retrieval) \cite{Zhang_2025_CVPR}. Candidate pages from sparse and dense branches are merged, with overlapping results removed to ensure unique page-level candidates (fusion and deduplication). For each page $p_i$ in the merged set, we compute scores from both branches and apply consistency-constrained fusion:

\begin{equation}
s(i) = \alpha s_{\text{sparse}}(i) + (1-\alpha)s_{\text{dense}}(i) - \lambda \Delta(i)
\end{equation}

where $s_{\text{sparse}}(i)$ and $s_{\text{dense}}(i)$ are sparse lexical and dense similarity scores, respectively, and $\Delta(i)$ penalizes cross-modal inconsistency when sparse and dense signals disagree. If confidence is low (e.g., entropy over fused scores, or missing required evidence type), VLD-RAG runs an additional retrieval step that refines $\mathcal{I}$ and issues new sub-queries (\textbf{iterative retrieval / agent loop})---following iterative reasoning agent patterns for visual document RAG \cite{wang2025vidorag}.

\begin{algorithm}[t]
\caption{Hybrid Retrieval with Modality-Consistent Fusion}
\label{alg:vldrag-retrieval}
\begin{algorithmic}[1]
\REQUIRE Query $q$, document $D = \{p_i\}_{i=1}^{N}$, visual encoder $f_{\text{vis}}$, visual embeddings $\{\mathbf{v}_i\}$, fusion weight $\alpha$, consistency penalty $\lambda$, retrieval depth $K$
\ENSURE Ranked evidence pages $\mathcal{P}_{\text{out}}$

\STATE Extract keywords: $k \leftarrow \text{KW}(q)$
\STATE Generate hypothetical document: $h \leftarrow \text{GenerateHypotheticalDoc}(q)$
\STATE $\mathbf{e}_h \leftarrow f_{\text{vis}}(h)$

\STATE $\mathcal{P}_{\text{sparse}} \leftarrow \text{SparseRetrieval}(k, D, K)$
\STATE $\mathcal{P}_{\text{dense}} \leftarrow \text{ANN-Search}(\mathbf{e}_h, \{\mathbf{v}_i\}, K)$

\STATE $\mathcal{P}_{\text{candidates}} \leftarrow \mathcal{P}_{\text{sparse}} \cup \mathcal{P}_{\text{dense}}$
\FOR{each page $p_i \in \mathcal{P}_{\text{candidates}}$}
    \STATE $\Delta(i) \leftarrow |s_{\text{sparse}}(i) - s_{\text{dense}}(i)|$
    \STATE $s(i) \leftarrow \alpha s_{\text{sparse}}(i) + (1-\alpha)s_{\text{dense}}(i) - \lambda \Delta(i)$
\ENDFOR

\STATE $\mathcal{P}_{\text{out}} \leftarrow \text{TopK}(\text{SortDescending}(\mathcal{P}_{\text{candidates}}, s), K)$
\STATE \textbf{return} $\mathcal{P}_{\text{out}}$
\end{algorithmic}
\end{algorithm}

\subsection{Evidence-Grounded Generation}

Retrieved evidence is represented as page IDs, optional region crops and parsed markdown spans (reconstructed from serialized markdown), and optional evidence type tags (table/chart/paragraph). A vision-language generator produces the answer conditioned on $q$ and $E$:
\begin{equation}
\hat{y} = g_{\phi}(q, E)
\end{equation}
where $g_{\phi}$ is a vision-language model parameterized by $\phi$ that generates the answer $\hat{y}$ based on the query $q$ and evidence $E$. The answer is then validated against the evidence:
\begin{equation}
\text{verify}(\hat{y}, E) \geq \tau
\end{equation}
where $\tau$ is a confidence threshold. If validation fails, the query is refined:
\begin{equation}
q' = R(q, E, r)
\end{equation}
where $R(\cdot)$ is a refinement function that generates a new query $q'$ based on the original query $q$, evidence $E$, and refinement signal $r$ (e.g., missing evidence or low confidence).

We pack evidence in a structured format (page thumbnails/crops + parsed markdown spans reconstructed from serialized content + metadata) to maximize grounding and minimize context length. The markdown serialization enables accurate reconstruction of tables and structured elements, which are then utilized in the retrieval and generation process. This is complementary to unified retrieval--generation trends for efficient long document understanding \cite{shi2026urag}.

\section{Experimentation}
\label{sec:experimentation}

\subsection{Datasets}

We evaluate our approach on two comprehensive multimodal document understanding benchmarks that cover diverse document types and tasks.

We use the LongDocURL benchmark \citep{deng2025longdocurl}, a comprehensive multimodal long document benchmark that integrates understanding, reasoning, and locating tasks. LongDocURL provides a standardized evaluation framework for long-document systems, featuring documents with up to 150 pages and covering various document sources including books, reports, manuals, project proposals, meeting minutes, and work summaries. The benchmark includes three main task categories: (1) \textbf{Understanding}: extracting information from documents by identifying keywords and parsing document structures, (2) \textbf{Reasoning}: processing numerical information through counting, calculating, comparing, and summarizing, and (3) \textbf{Locating}: analyzing relations among different types of elements and cross-element locating tasks. This benchmark enables us to assess our model's capabilities across multiple dimensions of long document understanding.

We evaluate on MMLongBench-Doc, which focuses on multimodal question answering over long documents with rich visual structure. The benchmark provides page-level annotations identifying evidence relevant to each question, enabling comprehensive evaluation of retrieval and generation capabilities.

\subsection{Experimental Setup}

Table~\ref{tab:appendix_dataset_statistics} presents detailed statistics of the datasets used in our experiments. Both benchmarks focus on multimodal question answering over long visually-rich documents, but differ in their characteristics and evaluation focus.

MMLongBench-Doc consists of 135 documents with an average of 47.5 pages per document, totaling 1,082 question-answer pairs. The dataset covers diverse document types (categorized as "Open") and contains an average of 21,214 tokens per document. Notably, 33.0\% of questions require evidence from multiple pages, and 22.6\% involve cross-element reasoning (e.g., relating information from tables, figures, and text). This benchmark emphasizes multimodal understanding and cross-page evidence retrieval.

LongDocURL is a larger-scale benchmark with 396 documents spanning 8 different document types, including books, reports, manuals, project proposals, meeting minutes, and work summaries. Documents are substantially longer, averaging 85.6 pages and 43,622.6 tokens per document. The benchmark contains 2,325 question-answer pairs, with queries averaging 35.5 tokens and reaching up to 277 tokens for complex questions. LongDocURL presents more challenging multi-page retrieval scenarios, with 52.9\% of questions requiring evidence from multiple pages and 37.1\% involving cross-element reasoning. The higher multi-page and cross-element percentages reflect the benchmark's focus on complex long-document understanding tasks that require reasoning across document structures and page boundaries.

For \textbf{hypothetical document generation}, we use Gemma3n \citep{gemma_3n_2025} (4B parameters) to generate query-conditioned hypothetical documents that approximate ideal answer-bearing pages, bridging the semantic gap between questions and document pages. \textbf{Vision-language embedding}: dense representations are computed using the ColPali vision encoder, which produces fixed-dimensional single-vector embeddings from page-level document images, capturing both visual layout and textual semantics. \textbf{Sparse retrieval}: keyword-based retrieval is configured with term frequency saturation $k_1=1.5$ and length normalization $b=0.75$, fixed across all experiments.

We evaluate retrieval performance using standard ranking metrics computed at the page level. \textbf{Recall@K} is the proportion of ground-truth evidence pages retrieved within top-$K$. \textbf{Precision@K} is the proportion of retrieved pages that are relevant. \textbf{MRR@K} is the Mean Reciprocal Rank of the first relevant page. \textbf{NDCG@K} is the Normalized Discounted Cumulative Gain, reflecting ranking quality with position-aware relevance.

Dense embeddings are indexed in a vector database supporting cosine similarity search. Vision-language encoding is performed on GPUs for efficiency. All evaluations are conducted under a \textbf{closed-document setting}, where retrieval is restricted to pages belonging to the same document as the query.

\subsection{Main Results}

We evaluate VLD-RAG on MMLongBench-Doc and LongDocURL benchmarks.

\begin{table*}[!htbp]
    \centering
    \caption{Retrieval performance evaluation across varying top-$K$ settings ($K \in \{1,3,5\}$) measured in percentage. Performance comparison of VLD-RAG against M3DocRAG \citep{Cho2024M3DocRAG}, MDocAgent (Text/Image), and MoLoRAG \citep{wu2025molorag} on MMLongBench-Doc and LongDocURL benchmarks. Our method demonstrates superior performance in Recall, NDCG, and MRR metrics across all $K$ values on both datasets; MoLoRAG achieves higher Precision scores at $K=3$ and $K=5$, whereas VLD-RAG's stronger Recall and ranking performance suggests enhanced effectiveness for retrieval-augmented generation tasks that prioritize comprehensive evidence page coverage.}
    \label{tab:retrieval_comp}
    \vspace{0.3em}
    \resizebox{0.82\linewidth}{!}{

     \begin{tabular}{cc|cccc|cccc} 
      \toprule
      \rowcolor{COLOR_MEAN}  & & \multicolumn{4}{c|}{MMLongBench} & \multicolumn{4}{c}{LongDocURL} \\
        \rowcolor{COLOR_MEAN} \multirow{-2}{*}{\textbf{Top-$K$}} & \multirow{-2}{*}{\textbf{Method}} & \textbf{Recall} & \textbf{Precision} & \textbf{NDCG} & \textbf{MRR}  & \textbf{Recall} & \textbf{Precision} & \textbf{NDCG} & \textbf{MRR}   \\ \midrule 
       \multirow{5}{*}{$1$} & M3DocRAG & 43.31 & 56.67 & 56.67 & 56.67 & 46.84 & 64.66 & 64.66 & 64.66 \\ 
        & MDocAgent (Text) & 29.30 & 38.99 & 38.99 & 38.99 & 42.03 & 58.37 & 58.37 & 58.37 \\ 
        & MDocAgent (Image) & 43.79 & 57.49 & 57.49 & 57.49 & 46.80 & 64.57 & 64.57 & 64.57 \\ 
        & MoLoRAG & 45.46 & 59.95 & 59.95 & 59.95 & 48.98 & \textbf{67.71} & \textbf{67.71} & \textbf{67.71} \\ 
       \rowcolor{orange!20} \cellcolor{white} & VLD-RAG (ours) & \textbf{48.92} & \textbf{64.52} & \textbf{64.52} & \textbf{64.52} & \textbf{50.72} & 65.22 & 65.22 & 65.22 \\ \midrule 

       \multirow{5}{*}{$3$} & M3DocRAG & 64.17 & 31.62 & 54.13 & 65.36 & 67.00 & 33.78 & 58.23 & 72.51  \\ 
       & MDocAgent (Text) & 43.21 & 20.77 & 37.13 & 45.26 & 58.53 & 29.33 & 54.12 & 65.28 \\ 
       & MDocAgent (Image) & 64.74 & 31.97 & 54.75 & 66.12 & 66.67 & 33.62 & 58.26 & 72.47 \\ 
        & MoLoRAG & 67.22 & \textbf{40.81} & 57.34 & 68.56 & 70.04 & \textbf{36.41} & 61.56 & 75.78 \\ 
       \rowcolor{orange!20} \cellcolor{white} & VLD-RAG (ours) & \textbf{70.16} & 35.48 & \textbf{68.83} & \textbf{72.58} & \textbf{76.09} & 34.78 & \textbf{70.75} & \textbf{74.64} \\ \midrule
       
        \multirow{5}{*}{$5$} & M3DocRAG & 72.00 & 22.58 & 54.06 & 66.92 & 74.32 & 23.34 & 58.05 & 73.83 \\ 
        & MDocAgent (Text) & 50.60 & 15.48 & 37.19 & 46.98 & 65.41 & 20.41 & 53.97 & 66.55 \\ 
        & MDocAgent (Image) & 71.45 & 22.37 & 54.58 & 67.53 & 74.60 & 23.50 & 58.06 & 73.90  \\ 
        & MoLoRAG & 74.13 & \textbf{35.83} & 57.29 & 69.63 & 77.14 & \textbf{26.13} & 61.30 & 76.88 \\ 
       \rowcolor{orange!20} \cellcolor{white} & VLD-RAG (ours) & \textbf{77.96} & 24.52 & \textbf{71.82} & \textbf{73.39} & \textbf{81.16} & 23.48 & \textbf{73.57} & \textbf{75.72} \\  \bottomrule
        
    \end{tabular}
    
    }
    
\end{table*}

Table~\ref{tab:retrieval_comp} presents retrieval performance comparison under top-$K$ settings ($K \in \{1, 3, 5\}$) across different methods.

The experimental results demonstrate consistent improvements of VLD-RAG over baseline methods across both benchmarks. On MMLongBench-Doc, VLD-RAG achieves the highest Recall@K across all $K$ settings, with substantial improvements over previous methods. Notably, VLD-RAG achieves the best NDCG and MRR scores at all $K$ values, with particularly notable gains in ranking quality metrics. While some baseline methods achieve higher Precision at intermediate $K$ values, VLD-RAG's superior Recall and ranking quality metrics (NDCG, MRR) indicate better overall retrieval effectiveness.

On LongDocURL, VLD-RAG shows particularly strong performance in Recall and ranking metrics. At $K=1$, VLD-RAG achieves strong Recall performance, though some baselines achieve higher Precision. The advantage becomes more pronounced at higher $K$ values, with substantial improvements in Recall metrics. VLD-RAG consistently achieves the best NDCG scores across all $K$ settings, with notable improvements over existing approaches. Similarly, MRR metrics demonstrate strong performance, outperforming all baselines.

\begin{figure}[!htbp]
    \centering
    \includegraphics[width=0.85\columnwidth]{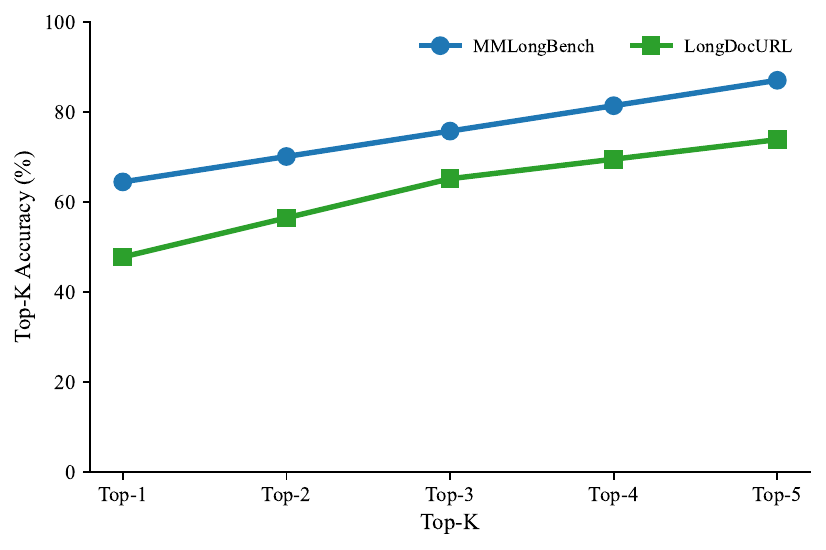}\\[0.6em]
    \includegraphics[width=0.85\columnwidth]{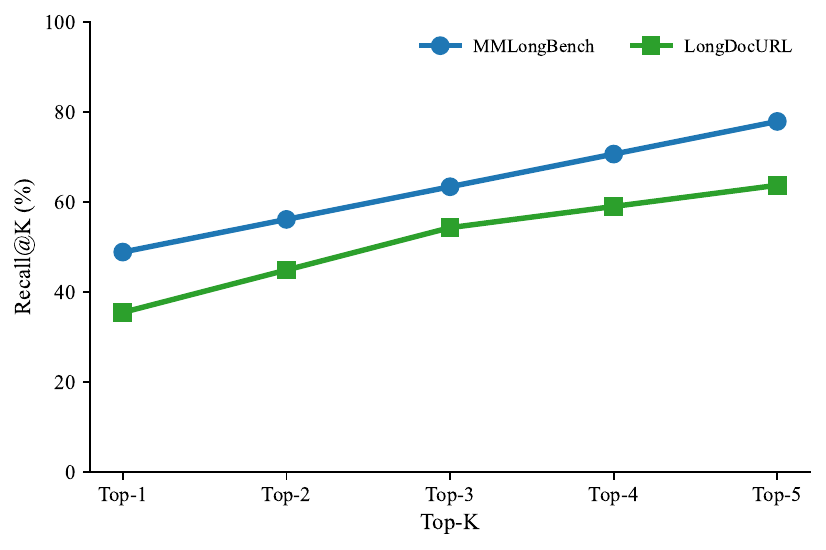}
    \caption{The effect of top-$K$ on retrieval performance across benchmarks. \textbf{Top:} Top-$K$ Accuracy (proportion of queries where at least one relevant page is retrieved within top-$K$). \textbf{Bottom:} Recall@$K$ (coverage of ground-truth evidence pages across retrieval depths). VLD-RAG (MMLongBench and LongDocURL) shows consistent gains as $K$ increases; the hybrid retrieval approach improves both accuracy and recall over the evaluated $K$ range, with MMLongBench-Doc achieving higher accuracy and LongDocURL showing strong recall trends.}
    \label{fig:topk_trends}
\end{figure}

The results reveal several important patterns. First, as $K$ increases from 1 to 5, Recall improves substantially for all methods, with VLD-RAG showing the largest absolute gains across both benchmarks. Second, LongDocURL shows generally higher Precision values, whereas NDCG improvements are lower compared to MMLongBench-Doc, suggesting different document complexity distributions. Third, MDocAgent \citep{han2025mdocagent} (Text) consistently underperforms, while MDocAgent (Image) and M3DocRAG \citep{Cho2024M3DocRAG} show competitive performance, yet fall short of VLD-RAG's hybrid approach. MoLoRAG \citep{wu2025molorag} demonstrates strong Precision performance at intermediate $K$ values, particularly on LongDocURL, but VLD-RAG's superior Recall and ranking metrics indicate better overall retrieval effectiveness. Although some baseline methods achieve strong Precision performance, VLD-RAG's superior Recall and ranking metrics demonstrate the effectiveness of our agentic verification and hybrid retrieval framework.

\begin{table*}[!htbp]
    \centering
    \caption{Dataset statistics for the datasets used in our experiments. MMLongBench-Doc and LongDocURL differ in scale (number of documents, pages, and QA pairs) and in the proportion of multi-page and cross-element questions; the table summarizes document count, type, average pages and tokens, QA count, and the percentage of questions requiring multi-page or cross-element reasoning, which reflects the benchmarks' focus on long visually-rich document understanding.}
    \label{tab:appendix_dataset_statistics}
    \vspace{0.3em}
    \resizebox{0.95\linewidth}{!}{
    \begin{tabular}{lccccccccc}
        \toprule
        \textbf{Dataset} & \textbf{\#Docs} & \textbf{Types} & \textbf{Avg Pages} & \textbf{Avg Tokens} & \textbf{\#QA} & \textbf{Avg Q Tokens} & \textbf{Max Q Tokens} & \textbf{Multi-page (\%)} & \textbf{Cross-element (\%)} \\
        \midrule
        MMLongBench-Doc & 135 & Open & 47.5 & 21,214.1 & 1,082 & \textemdash{} & \textemdash{} & 33.0 & 22.6 \\
        LongDocURL & 396 & 8 & 85.6 & 43,622.6 & 2,325 & 35.5 & 277 & 52.9 & 37.1 \\
        \bottomrule
    \end{tabular}
    
    }
    
\end{table*}

\subsection{Analysis}

Our experimental results demonstrate the effectiveness of the hybrid retrieval framework. The combination of sparse lexical retrieval (for keyword matching) and vision embeddings (for semantic understanding) provides complementary strengths, resulting in robust retrieval across diverse question types and document structures. VLD-RAG consistently outperforms single-method baselines across both benchmarks. The substantial improvements in NDCG over previous methods, particularly on LongDocURL, indicate that our hybrid approach better captures position-aware relevance, which is crucial for multi-page evidence retrieval where the order of retrieved pages matters for downstream generation.

On MMLongBench-Doc, our method achieves strong performance in Recall, NDCG, and MRR metrics across different top-$K$ settings, with particularly notable gains in ranking quality metrics. On LongDocURL, VLD-RAG demonstrates superior retrieval quality, particularly in Recall and NDCG metrics, highlighting the framework's ability to handle heterogeneous page layouts and long-form documents. The larger performance gaps on LongDocURL compared to MMLongBench-Doc suggest that our approach is especially effective for documents with more complex structures and longer page sequences.

The integration of hypothetical document generation helps bridge the semantic gap between queries and document pages, particularly when evidence is embedded within complex layouts, tables, or figures. This contributes to improved dense retrieval performance compared to direct query encoding. The consistent improvements in Recall metrics across all $K$ settings, especially at higher $K$ values where more candidate pages are considered, demonstrate that hypothetical documents enable better semantic matching even when exact keyword matches are sparse.

While some baseline methods achieve higher Precision at intermediate $K$ values on both benchmarks, VLD-RAG's superior Recall and ranking metrics (NDCG, MRR) indicate a better balance for retrieval-augmented generation tasks. Higher Recall ensures that relevant evidence pages are not missed, which is critical for comprehensive answer generation, while superior NDCG and MRR reflect better ranking quality that prioritizes the most relevant pages earlier in the retrieval list.

\section{Conclusion}
\label{sec:conclusion}

We presented VLD-RAG, an agentic vision--language retrieval-augmented generation framework for long and visually-rich documents. By combining a training-free dual-index architecture (vision embeddings and text resources from parsed markdown) and modality-consistent fusion with iterative retrieval, VLD-RAG improves both retrieval robustness and end-to-end question answering over multi-page visual documents. Our approach addresses key limitations of existing RAG systems by preserving visual layout information through page-level embeddings while maintaining precise lexical matching capabilities through structured markdown parsing. The verifier-guided agentic retrieval loop enables reliable multi-page evidence discovery, demonstrating significant improvements in Recall, NDCG, and MRR metrics across different top-$K$ settings.

Experiments on LongDocURL and MMLongBench-Doc show consistent gains over text-only and vision-only baselines, with particularly notable improvements in ranking quality metrics that are crucial for retrieval-augmented generation scenarios. The hybrid retrieval strategy effectively combines the complementary strengths of sparse lexical search and dense semantic retrieval, while the modality-consistency constraints help mitigate cross-modal disagreements that can degrade retrieval quality. Future work may extend the framework to more modalities (e.g., audio, video), explore adaptive fusion strategies that dynamically weight modalities based on query characteristics, and scale to even larger document corpora with more efficient indexing and retrieval mechanisms.

\bibliography{main}
\bibliographystyle{paperstyle}

\newpage
\onecolumn
\appendix
\section{Instructions}
\label{sec:appendix_instructions}

\begin{figure*}[htbp]
    \centering
    \includegraphics[width=0.85\textwidth]{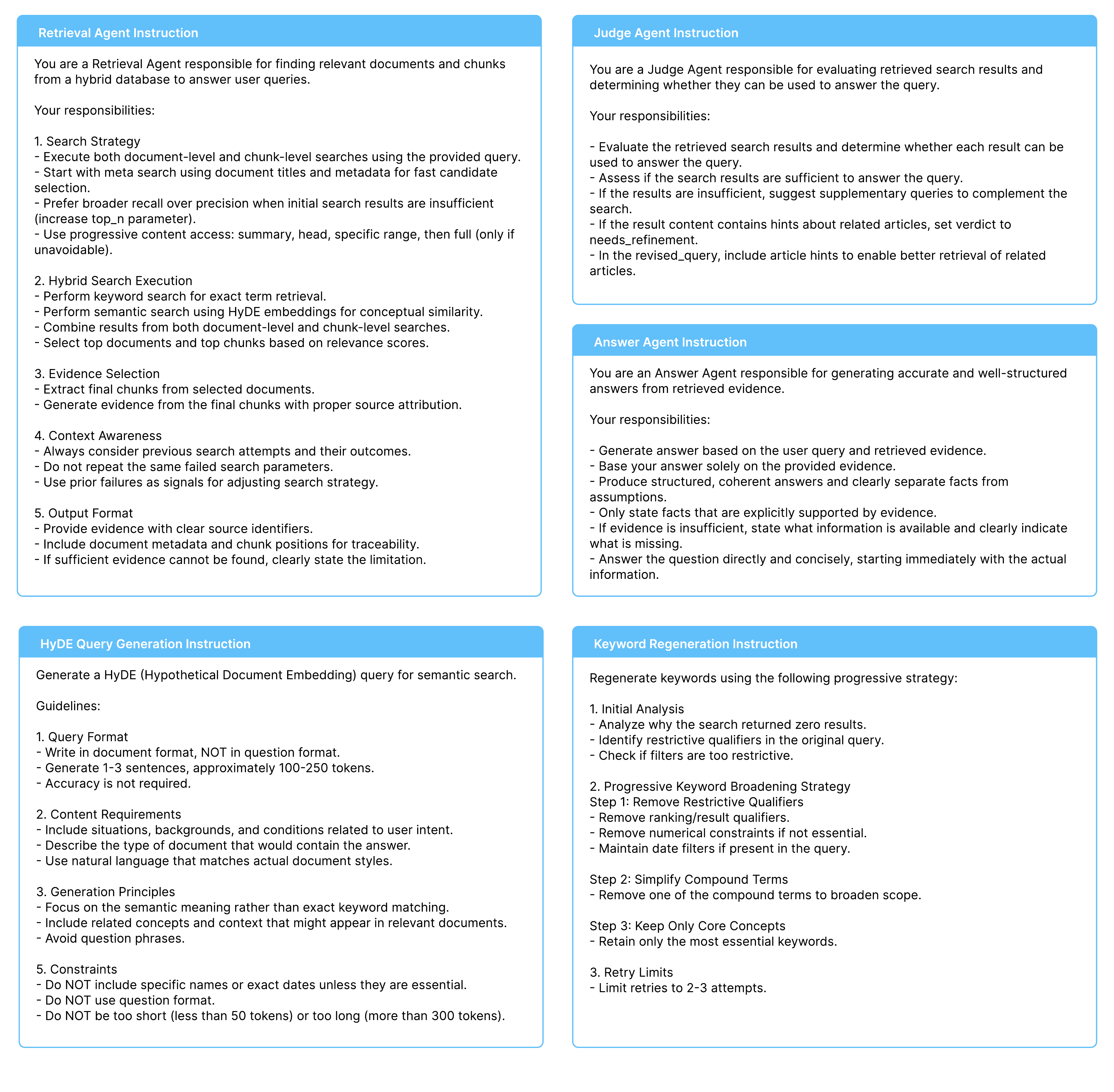}
    \caption{Instructions for the VLD-RAG system, defining agent roles, workflows, and coordination protocols for retrieval, answer generation, and validation. The instructions specify how the Retrieval Agent, Answer Agent, and Validation Agent interact; they also describe the iterative refinement process (e.g., query refinement when evidence is missing or validation fails), enabling reproducible agentic behavior across experiments.}
    \label{fig:instructions}
\end{figure*}

Figure~\ref{fig:instructions} presents the detailed instructions used in the VLD-RAG system for agent coordination and task execution. These instructions define the roles and responsibilities of each agent component in the retrieval-augmented generation pipeline.

The instructions specify the workflow for the \textbf{Retrieval Agent}, which is responsible for selecting relevant document pages and evidence chunks based on the query and retrieval intents. The \textbf{Answer Agent} receives the selected evidence and generates comprehensive answers grounded in the retrieved information. The \textbf{Validation Agent} verifies the generated answers against the evidence, checking for factual consistency, completeness, and proper citation of sources.

The instructions also define the iterative refinement process, where queries are refined based on validation feedback and missing evidence detection. This agentic coordination enables robust multi-hop reasoning and ensures that answers are well-grounded in the retrieved evidence across multiple pages and document elements.

\end{document}